\documentclass[aps,prd,twocolumn,letterpaper,superscriptaddress, showpacs, amsmath, amssymb]{revtex4}

\usepackage{graphicx}
\usepackage{amsmath,amssymb}

\begin{document}

%%%%%%%%%%%%%%%%%%%%%%%%%%%%%%%%%%%%%%%%%%%%%%%%%%
%%%%%%%%%%%%%%%%%%%%%%%%%%%%%%%%%%%%%%%%%%%%%%%%%%

\title{Quasithermal neutrinos from rotating protoneutron stars\\
born during core collapse of massive stars}

\affiliation{Hubble Fellow --- Institute for Advanced Study, Princeton, New Jersey 08540, USA}
\affiliation{International Centre for Theoretical Physics, Trieste 34014, Italy}
\affiliation{CCAPP, The Ohio State University, Columbus, Ohio 43210, USA}
\affiliation{Department of Astronomy, The Ohio State University, Columbus, Ohio 43210, USA}

\author{Kohta Murase$^{1,3}$, Basudeb Dasgupta$^{2,3}$, and Todd A. Thompson$^{3,4}$}

\date{\today}

\begin{abstract}
Rotating and magnetized protoneutron stars (PNSs) may drive relativistic magneto-centrifugally accelerated winds as they cool immediately after core collapse.  The wind fluid near the star is composed of neutrons and protons, and the neutrons become relativistic while collisionally coupled with the ions.  Here, we argue that the neutrons in the flow eventually undergo inelastic collisions around the termination shock inside the stellar material, producing $\sim0.1-1$~GeV neutrinos, without relying on cosmic-ray acceleration mechanisms.  Even higher-energy neutrinos may be produced via particle acceleration mechanisms.  We show that PINGU and Hyper-Kamiokande can detect such neutrinos from nearby core-collapse supernovae, by reducing the atmospheric neutrino background via coincident detection of MeV neutrinos or gravitational waves and optical observations.  Detection of these GeV and/or higher-energy neutrinos would provide important clues to the physics of magnetic acceleration, nucleosynthesis, the relation between supernovae and gamma-ray bursts, and the properties of newly born neutron stars.  
\end{abstract}

\pacs{95.85.Ry, 97.60.Bw, 98.70.Rz \vspace{-0.3cm}}
% 95.85.Ry Neutrino, muon, pion, and other elementary particles; cosmic rays

\maketitle

%%%%%%%%%%%%%%%%%%%%%%%%%%%%%%%%%%%%%%%%%%%%%%%%%%
%%%%%%%%%%%%%%%%%%%%%%%%%%%%%%%%%%%%%%%%%%%%%%%%%%

\section{Introduction}
Protoneutron stars (PNSs) are produced in core-collapse supernovae (CCSNe), and cool via radiation of MeV neutrinos on a time scale of $\sim10-100$~s~\citep[e.g.,][]{bl86,pon+99}.  A fraction of these thermal neutrinos deposit their energy in the PNS atmosphere, driving a wind with mass loss rate $\dot{M}$ that injects energy into the shocked stellar material and forms a PNS wind driven bubble in the SN cavity~\citep[e.g.,][]{woo+94}.  For nonrotating and nonmagnetic PNSs, the wind kinetic energy is tiny compared to the CCSN explosion energy, and the wind is nonrelativistic throughout the cooling epoch~\citep[e.g.,][]{qw96}.  However, if PNSs are rotating and magnetized, they transit from nonrelativistic and thermally driven to relativistic and Poynting dominated winds~\citep[e.g.,][]{tho+04,met+11}.  The magnetization of the flow is~\cite{tho+04,met+11,mic69}
\begin{equation}
\sigma\approx\frac{B_{\rm dip}^2R_{\rm ns}^4\Omega^2}{\dot{M} c^3}f_{\rm op}^2, 
\end{equation}
where $R_{\rm ns}$ is the PNS radius, $B_{\rm dip}$ is the surface dipolar field strength, $\Omega$ is the angular frequency of the PNS, and $f_{\rm op}$ takes into account that the outflow comes from only the open fraction of the PNS surface.  The transition from nonrelativistic to relativistic winds occurs at $\sigma\sim1$ where the Alfv\'en speed becomes $\sim c$ as $\dot{M}$ decreases.

Neutrons are naturally loaded in the PNS wind via neutrino heating, and they are also important for synthesis of heavy nuclei~\cite{rob+10,met+11b}.  Around the base of the outflow, neutrons and ions are tightly coupled via elastic collisions, so neutrons are accelerated together with ions as the Poynting-dominated outflow is accelerated by magnetic fields.  Although it has been known that the magnetic energy of the flow is not efficiently converted to kinetic energy in ideal MHD~\cite{gj70}, efficient acceleration is strongly motivated to explain the jets of gamma-ray bursts (GRBs) and active galaxies, and can be achieved by rapid time-variability or magnetic dissipation in the flow~\cite[see a review][and references therein]{kom11}.  Once the outflow becomes relativistic enough to exceed the pion-production threshold, inelastic collisions are naturally expected as the main dissipation process of relativistic neutrons.  The relevance of the $np$ reaction has been suggested in the context of GRBs~\cite{neutron}, where internal collisions due to outflow inhomogeneities or neutron decoupling have been considered.  However, this process is even more generic.  As we show in this work, as rotating magnetized PNSs cool and $\dot{M}$ decreases, neutron decoupling occurs at radii where the flow has already become relativistic and where the neutrons are above the pion production threshold.  The neutrons then interact with the material decelerated by the shock and possibly with the overlying stellar material, producing $0.1-1$~GeV neutrinos.  Detecting this signal would probe the otherwise completely obscured process of jet acceleration and the physics of rotating and magnetized PNS birth during the core collapse of massive stars. 

Rapidly rotating and/or strongly magnetized PNSs are particularly interesting objects since they may be related to various explosion phenomena.  Estimates suggest that $\gtrsim10$\% of CCSNe lead to magnetars with $B_{\rm dip}\sim{10}^{14-15}$~G~\cite{magrev}, which may be generated by rapid rotation via the dynamo mechanism~\cite{dt92}.  Rotation and/or magnetization can also modify the explosion dynamics.  If the wind power exceeds $\sim{10}^{48}~{\rm erg}~{\rm s}^{-1}$, the wind can be collimated, forming relativistic jets~\cite{buc+07,kb07} that may lead to GRBs~\cite{uso94,tho+04,met+11}.  Although only a fraction (up to a few percent) of CCSNe harboring relativistic jets may be observed as CCSNe with a relativistic component~\cite{ccsnjet}, trans-relativistic CCSNe~\cite{sod+06,sod+10} may belong to such a class.  If the rotation rate or magnetic field strength is not sufficiently high, a quasispherical wind or a pair of choked jets hidden by the stellar material may result~\cite{wm03}, potentially leading to (nonrelativistic) hypernovae and super-luminous supernovae~\cite{tho+04,met+11,whe+00}.  On the theoretical side, magnetically driven CCSNe have been of interest and studied for many years~\cite{moscow,mh79}.  

The organization of this paper is as follows.  First, we consider a neutrino driven PNS wind, by which baryons are naturally loaded in the outflow.  As a PNS cools, the outflow becomes baryon poor and $\sigma$ becomes $\gg1$, so it will be magnetically accelerated rather than thermally accelerated.  We show that the transition can happen in the Kelvin-Helmholtz cooling time scale when the PNS is rotating and magnetized.  In Section IV, we show that neutron decoupling happens after neutrons are accelerated together with ions while they are coupled.  Additional neutrons may be generated by photodisintegration of nuclei, if electrons are accelerated around the termination shock of the PNS wind.  Then, in Section VI, we show that $\sim0.1-1$~GeV neutrinos should be generated via the $np$ reaction and further boosts to the expected flux may come from the neutron-proton-converter (NPC) acceleration~\cite{npc} and/or shock acceleration mechanisms.  In Section VII, we see that these neutrinos can be detected by planned facilities such as PINGU~\cite{pingu} and Hyper-Kamiokande (HK)~\cite{hk}.  Finally, as discussed in Section VIII, their characterization would allow us to probe the physics of PNSs, early dynamics of the jet or wind, and magnetic acceleration in an environment inaccessible to photons. 

Throughout this work, we use $Q_x\equiv Q/{10}^x$ in cgs units unless otherwise specified.  

%%%%%%%%%%%%%%%%%%%%%%%%%%%%%%%%%%%%%%%%%%%%%%%%%%
%%%%%%%%%%%%%%%%%%%%%%%%%%%%%%%%%%%%%%%%%%%%%%%%%%

\section{Baryon loading by neutrino driven winds}
Mass loss from PNSs occurs during the PNS cooling phase by neutrino heating, mainly via $\nu_en\rightleftharpoons e^-p$ and $\bar{\nu}_ep\rightleftharpoons e^+n$.  In unmagnetized winds~\cite{qw96,met+11} 
\begin{eqnarray}
\dot{M}_{\nu}&\approx&1.4\times{10}^{-4} M_{\odot}~{\rm s}^{-1}~L_{\nu,52}^{5/3}{(\varepsilon_{\nu}/15~{\rm MeV})}^{10/3}\nonumber\\
&\times&{(1+\epsilon_{\rm es})}^{5/3}R_{\rm ns,6}^{5/3}{(M_{\rm ns}/1.4~M_\odot)}^{-2},
\end{eqnarray} 
where $L_\nu$ is the neutrino ($\nu_e+\bar{\nu}_e$) luminosity and $\varepsilon_\nu$ is the typical neutrino energy, which can be affected by rotation~\cite{tho+05}.  Additional heating due to inelastic electron scattering gives a correction $f_{\rm es}\equiv1+\epsilon_{\rm es}$.  The existence of magnetic fields modifies $\dot{M}$ from the above expression in three ways.  Firstly, the mass-loss rate is reduced by $f_{\rm op}$ since only the open fraction of the PNS surface contributes to the outflow.  For times much less than time at which $\sigma$ becomes equal to unity ($t\ll t_{\rm tr}$), $f_{\rm op}\sim1$ is expected while it becomes smaller than unity at later times.  If we assume $R_Y/R_{\rm lc}\sim{\rm min}[1,0.3\sigma^{0.15}]$ for $R_Y>R_{\rm ns}$ (where $R_Y$ is the Y point radius where the close zone ends in the magnetic equatorial plane and $R_{\rm lc}\equiv c/\Omega$ is the light cylinder radius), following Ref.~\cite{met+11}, we obtain $f_{\rm op}=(1-\cos \theta_{\rm op})\sim0.07-0.14$ for $\sigma\sim1-100$ and $P=0.01$~s, using $\theta_{\rm op}\sim2{\sin}^{-1}(\sqrt{R_{\rm ns}/R_Y})$.  Secondly, $\dot{M}$ is enhanced by $f_{\rm cen}$ due to the centrifugal force~\cite{met+08}.  Magnetic fields and rotation are so strong and fast that the centrifugal force increases the scale height in the heating region, which can lead to $f_{\rm cen}\gtrsim1$.  The third effect comes from the fact that electrons and positrons participating in the charged-particle reactions are restricted to discrete Laundau levels, but this is negligible for $B_{\rm dip}\lesssim{10}^{16.5}$~G~\cite{dq04}.  Taking into account these effects of magnetic fields, we describe the baryon mass-loss rate by $\dot{M}_b\approx\dot{M}=\dot{M}_{\nu}f_{\rm op}f_{\rm cen}$~\cite{met+11}.  In this work, we simply regard $f_{\rm op}$ and $f_{\rm cen}$ as prefactors, since their evolution is uncertain.  By setting $\sigma\sim1$ in Eq.~(1), we find that the transition to relativistic flow ($t=t_{tr}$) occurs when $\dot{M}_b\approx\dot{M}_{\rm tr}\simeq7.4\times{10}^{-8}M_{\odot}~{\rm s}^{-1}~B_{\rm dip,15}^2P_{-2}^{-2}f_{\rm op,-1}^2R_{\rm ns,6}^4$, where $P=2\pi/\Omega$.  

The neutrino luminosity decreases gradually as a power law, until the PNS becomes transparent to neutrinos at $t_{\rm thin}\sim10-100$~s~\cite{pon+99,met+11}. For $t>t_{\rm thin}$, $L_\nu$, $\varepsilon_\nu$ and $\dot{M}_\nu$ should decline rapidly.  Eventually, neutrino heating produces no mass loss, and instead $\dot{M}$ is described by the Goldreich-Julian density: $\dot{M}_{\rm GJ}\approx2.5\times{10}^{-17}M_{\odot}~{\rm s}^{-1}~\mu_{\pm,6}B_{\rm dip,15}P_{-2}^{-2}R_{\rm ns,6}^{3}$~\cite{gj69}, where $\mu_\pm$ is the pair-multiplicity.  In this late phase, we expect $\dot{M}_{b}\approx\dot{M}_{\rm GJ}m_p/(\mu_{\pm}m_eY_e)$, where $Y_e$ is the electron fraction.  Note that in this work we focus on low-entropy winds, as in the protomagnetar model of GRBs~\cite{met+11,met+11b}, that considers a Poynting-dominated jet driven by a central rapidly rotating magnetar.  Hence, pairs loaded at the base of the flow are irrelevant at the neutrino production site until $t\gtrsim t_{\rm thin}$, unlike the fireball model of GRBs.       

%%%%%%%%%%%%%%%%%%%%%%%%%%%%%%%%%%%%%%%%%%%%%%%%%%
%%%%%%%%%%%%%%%%%%%%%%%%%%%%%%%%%%%%%%%%%%%%%%%%%%

\section{Transition to magnetic field driven acceleration}
For the purpose of simple estimates, we hereafter assume $M_{\rm ns}=1.4M_\odot$, $R_{\rm ns}={10}^6$~cm, $L_\nu\propto t^{-1}e^{-t/t_{\rm thin}}$, $\varepsilon_\nu\propto L_\nu^{1/4}$~\cite{pon+99,met+11}.  Then, the transition time $t_{\rm tr}$, when the PNS wind becomes relativistic ($\sigma\sim1$) is
\begin{eqnarray}
t_{\rm tr}&\sim&8.0~{\rm s}~B_{\rm dip,15}^{-4/5}P_{-2}^{4/5}f_{\rm op,-1}^{-2/5} f_{\rm cen}^{2/5}\nonumber\\
&\times&L_{\nu 0,52}^{2/3}{(\varepsilon_{\nu 0}/15~{\rm MeV})}^{4/3}f_{\rm es}^{2/3},
\end{eqnarray}
where $L_{\nu 0}$ and $\varepsilon_{\nu 0}$ are defined at 1~s.  Therefore, if magnetic fields are strong and/or rotation is rapid, the PNS wind becomes relativistic 
at $t \ll t_{\rm thin}$.  If not, it will become so at $\sim t_{\rm thin}$ as $\dot{M}$ rapidly declines.  

The $\sigma$ parameter at $t\ll t_{\rm thin}$ is 
\begin{eqnarray}
\sigma(t)&\sim& 30 B_{\rm dip,15}^2P_{-2}^{-2}f_{\rm op,-1}f_{\rm cen}^{-1}L_{\nu 0,52}^{-5/3}\nonumber\\ 
&\times&{(\varepsilon_{\nu 0}/15~{\rm MeV})}^{-10/3}f_{\rm es}^{-5/3}t_{1.5}^{5/2},
\end{eqnarray}
which rapidly increases with time.  The $\sigma$ parameter at $t_{\rm thin}$ becomes
\begin{eqnarray}
\sigma(t_{\rm thin})&\sim&1100B_{\rm dip,15}^2P_{-2}^{-2}f_{\rm op,-1}f_{\rm cen}^{-1}L_{\nu 0,52}^{-5/3}\nonumber\\ 
&\times&{(\varepsilon_{\nu 0}/15~{\rm MeV})}^{-10/3}f_{\rm es}^{-5/3},
%{(t_{\rm thin}/50~{\rm s})}^{5/2}, 
\end{eqnarray}
and $\dot{M} (t_{\rm thin})\sim6.5\times{10}^{-11}~M_\odot~{\rm s}^{-1}~f_{\rm op,-1}f_{\rm cen}L_{\nu 0,52}^{5/3}\\{(\varepsilon_{\nu 0}/15~{\rm MeV})}^{10/3}f_{\rm es}^{5/3}$ for $t_{\rm thin}=50~{\rm s}$.  We focus on the epoch $t\lesssim t_{\rm thin}$, since the flux of relativistic neutrons producing quasithermal neutrinos decreases strongly for $t\gg t_{\rm thin}$ even though the wind becomes more relativistic. 

At $t\gtrsim t_{\rm tr}$, the PNS wind is accelerated mainly magnetically.  When the maximum Lorentz factor $\Gamma_{\rm max}\approx\sigma$ is achieved at the saturation radius $R_{\rm sat}\approx R_{\rm mag}$, we can parametrize $\Gamma(r)$ as~\cite{spr+03}  
\begin{equation}
\Gamma(r)\approx{\rm min}\left[\sigma,\sigma{\left(\frac{r}{R_{\rm mag}}\right)}^{1/3}\right],
\end{equation}
where $R_{\rm mag}\approx\pi c\sigma^2/(3\epsilon_{\rm rec}\Omega)\simeq5.0\times{10}^{15}~{\rm cm}~\sigma_3^2P_{-2}\epsilon_{\rm rec,-2}^{-1}$~\cite{spr+03,met+11}, and $\epsilon_{\rm rec}$ is a parameter characterizing magnetic reconnection~\cite{spr+03,ber13}.  Note that more efficient dissipation or radiative acceleration leads to indices larger than $1/3$ and smaller $R_{\rm sat}$.  Hence, our results below are relatively conservative.  

The density in the initial cavity left by the CCSN shock is so small that a PNS wind freely expands and easily sweeps through the cavity.  As soon as the wind hits the high-density CCSN ejecta, it is forced to slow down to a speed of order of the CCSN shock velocity.  Then, a hot magnetized subsonic bubble forms and its evolution depends on $\sigma$ and on the spin-down power of the PNS.  When the wind power is not high enough, a PNS wind driven bubble, which terminates at the reverse shock caused by the interaction between the flow and shocked stellar material, would be quasispherical~\cite{buc+07}.  If the wind power exceeds $\sim10^{48}~{\rm erg}~{\rm s}^{-1}$, depending on progenitor properties, the anisotropic thermal pressure can redirect the equatorial wind~\cite{buc+07}.  Then, the collimation happens at ${\rm a~few}\times{10}^{8}$~cm and bipolar flows are launched at a speed of $c$, which may lead to GRBs if the jets are successful in punching through the overlying progenitor star.   
In the jet case, additional baryon loading could occur around the jet-star boundary due to e.g., Kelvin-Helmholtz instabilities.  Although it can affect $\dot{M}$ around the sheath region, our concept can still be applied.  Note that the baryon loading is not necessarily a decisive parameter to make successful jets.  To see whether jets are finally successful and differentiate the fate of jet-driven explosions, parameters of the total jet power, jet duration, initial opening angle and density profile of progenitors are also important.  

We define the wind termination radius to be $R_w$.  In the case of a low-power quasispherical flow or a high-power biconical pair of jets, $R_w$ is regarded as the radius of the reverse shock caused by the interaction between the flow and preshocked stellar material.  When $R_w$ is smaller than $R_{\rm mag}$, the final Lorentz factor of the flow is
\begin{equation}
\Gamma (R_{w})\simeq13\sigma_{3}^{1/3}P_{-2}^{-1/3}\epsilon_{\rm rec,-2}^{1/3}R_{w,10}^{1/3}.
\end{equation}
The maximum Lorentz factor ($\sigma$; Eq.~6) is achieved only if there is no boundary to the flow so that the wind radius reaches $R_{w}\gtrsim R_{\rm mag}$, as could be achieved in a high-power wind that launches successful jets~\cite{buc+07,kb07} or if the acceleration happens more rapidly than assumed in Eq.~(6) (e.g., more efficient magnetic dissipation).
%(e.g., choked jets in the fireball scenario for GRBs). 

%%%%%%%%%%%%%%%%%%%%%%%%%%%%%%%%%%%%%%%%%%%%%%%%%%
%%%%%%%%%%%%%%%%%%%%%%%%%%%%%%%%%%%%%%%%%%%%%%%%%%

\section{Neutron decoupling}
Neutrons and ions have the same outflow velocity as long as they are coupled with $\langle\sigma_{\rm el}v\rangle\approx\sigma_{np}c$~\cite{neutron}.  Here $\sigma_{\rm el}$ is the elastic cross section and $\sigma_{np}\approx3\times{10}^{-26}~{\rm cm}^2$ is the inelastic cross section.  However, if the neutrons are decoupled during the flow acceleration, then they will have a smaller Lorentz factor than the ions.  Using the nucleon density $n_w=\dot{M_b}/(4\pi r^2 c m_p\Gamma)$ and Eq.~(6), $\tau_{np}\approx n_w \sigma_{np}(R_{\rm dec}/\Gamma)=1$ gives the decoupling radius of the neutrons:
\begin{equation}
R_{\rm dec}\simeq3.9\times{10}^8~{\rm cm}~\sigma_{3}^{-1}B_{\rm dip,15}^{6/5}P_{-2}^{-4/5}f_{\rm op,-1}^{6/5}\epsilon_{\rm rec,-2}^{-2/5}.   
\end{equation}
Initially, $R_{\rm dec}$ is larger than $R_{\rm mag}$ and $R_w$.  It becomes smaller as time and crosses $R_{\rm mag}$ and $R_w$.  The decoupling in the acceleration phase usually occurs at $\sigma(t)\ll\sigma(t_{\rm thin})$, i.e., $t\ll t_{\rm thin}$.  For $R_{\rm dec}<R_{w},R_{\rm mag}$, we expect the neutron flow to
have a Lorentz factor $\Gamma_n$ of
\begin{equation}
\Gamma(R_{\rm dec})\approx4.3B_{\rm dip,15}^{2/5}P_{-2}^{-3/5}f_{\rm
op,-1}^{2/5}\epsilon_{\rm rec,-2}^{1/5}.
\end{equation}

Because $R_{\rm mag}\propto\sigma^2$, $R_{\rm mag}$ should increase rapidly in time (Eqs.~1 and 2).  $R_w$ also increases, but more slowly.  Hence, if $R_{\rm mag}(t_{\rm tr})<R_w(t_{\rm tr})$ at $t_{\rm tr}$, then $R_{\rm mag}$ eventually overtakes $R_{w}$.  On the other hand, $R_{\rm dec}$ decreases with time, so it crosses $R_{\rm mag}$ and $R_w$ as long as $R_{\rm dec} (t_{\rm tr})$ is large enough (that is satisfied in our cases).  We define $t_a$ by $R_{\rm dec}(t_a)=R_{\rm mag}(t_a)$ and $t_b$ by $R_{\rm dec}(t_b)=R_w(t_b)$, respectively (see Fig.~1).  (Note that $t_a\lesssim t_{\rm thin}$ and $t_b\lesssim t_{\rm thin}$ since $\sigma$ abruptly increases around $t_{\rm thin}$ because of the rapid decrease in $\dot{M}$.)

Neutrons and ions achieve the same final Lorentz factor in the early phase ($t<{\rm max}[t_{a},t_{b}]$), whereas neutrons have lower final Lorentz factor in the later phase because of decoupling.  If $t_a<t_b$, we have $\Gamma_n\approx\Gamma(R_w)$ at $t<t_b$ and  $\Gamma_n\approx\Gamma(R_{\rm dec})$ at $t_b<t$.  If $t_b<t_a$, we obtain $\Gamma_n\approx\Gamma(R_{\rm mag})$ at $t<t_a$ and  $\Gamma_n\approx\Gamma(R_{\rm dec})$ at $t_a<t$.  If $t_a\ll t_{\rm thin}$, we have
\begin{eqnarray}
t_{a}&\sim&15~{\rm s}~B_{\rm dip,15}^{-16/25}P_{-2}^{14/25}f_{\rm op,-1}^{-6/25}f_{\rm cen}^{2/5}\epsilon_{\rm rec,-2}^{2/25}\nonumber\\
&\times&L_{\nu 0, 52}^{2/3}{(\varepsilon_{\nu 0}/15~{\rm MeV})}^{4/3}f_{\rm es}^{2/3},
\end{eqnarray}
using Eqs.~(1) and (2).  Note that the corresponding decoupling radius is order of ${10}^{10}$~cm for our fiducial parameters.  If $t_b\ll t_{\rm thin}$, assuming the typical velocity $V\approx R_w/t$, $t_b$ is estimated to be
\begin{eqnarray}
t_{b}&\sim&34~{\rm s}~B_{\rm dip,15}^{-8/35}P_{-2}^{12/35}f_{\rm op,-1}^{2/35}f_{\rm cen}^{2/7}\epsilon_{\rm rec,-2}^{-4/35}\nonumber\\
&\times&L_{\nu 0, 52}^{10/21}{(\varepsilon_{\nu 0}/15~{\rm MeV})}^{20/21}f_{\rm es}^{10/21}V_{8.5}^{-2/7}.
\end{eqnarray} 
Neutrons decay with proper lifetime of $886.7$~s, so the decay radius $R_\beta\approx2.7\times {10}^{14}~{\rm cm}~\Gamma_{n,1}$ is much longer than $R_{\rm dec}$, $R_{\rm mag}$ and $R_w$, in which we are interested. 

From Eqs.~(10) and (11), for a quasispherical bubble that nonrelativistically expands, we usually expect $t_a<t_b$.  However, if jets form and $R_w$ relativistically expands, $t_b<t_a$ is possible.  The decoupling radius at $t_a$ is of order ${10}^{10}$~cm, which implies that neutrons and ions are tightly coupled at the radius where the equatorial wind is redirected into a jet-like configuration.  Once the configuration is jet-like, we expect the resulting neutrino emission to be beamed along this axis, with consequences for the predicted fluence (see below). 

%%%%%%%%%%%%%%%%%%%%%%%%%%%%%%%%%%
\begin{figure}[t]
\includegraphics[width=0.75\linewidth,angle=270]{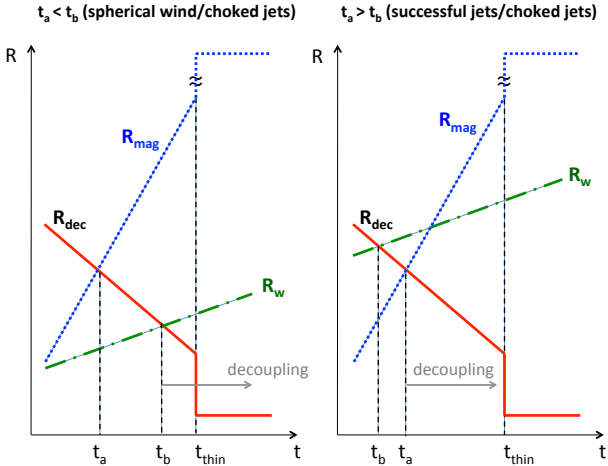}
\caption{Schematic picture of $R_{\rm mag}\propto\sigma^2$, $R_{\rm dec}\propto\sigma^{-1}$ and $R_{w}\propto t^{1+\chi}$.  In the case of a near-spherical wind, we typically expect $t_a<t_b$, since a magnetized bubble expands nonrelativistically and $\dot{M}$ is given through Eq.~(2).  In the case of jets, $\dot{M}$ could be different due to additional baryon loading, although our qualitative picture will not change.}
\vspace{-1.\baselineskip}
\end{figure}
%%%%%%%%%%%%%%%%%%%%%%%%%%%%%%%%%%

\section{Neutrons from disintegrated nuclei}
Possibly, neutrons may be produced within $R_w$ via disintegration of nuclei, as well as those not used up in nucleosynthesis.  Fe-group elements are mainly created if the initial electron fraction $Y_{e0}\gtrsim0.5$, while the neutron capture channel is dominant and $A>56$ nuclei are synthesized if $Y_{e0}\lesssim0.5$.  
%They could be seeds of ultra-high-energy cosmic rays (UHECRs) if they could survive (e.g., after successful jets break out)
Recently, nucleosynthesis of heavy nuclei have been studied in more detail in the context of PNS winds~\cite{rob+10,met+11b}.   

Nuclei may be spalled, which is relevant at $R_w\lesssim R_{\rm dec}$ since shocked nuclei collide with neutrons~\citep[e.g.,][and references therein]{hor+12}.  They can also be disintegrated by photons mainly via the giant-dipole resonance $A\gamma\rightarrow A' N$ if $2\Gamma\varepsilon\geq8.5~{\rm MeV}~{(A/56)}^{-1/6}$, where $\varepsilon$ is the target photon energy in the stellar frame~\cite{rac96}.  The corresponding threshold energy is $\varepsilon_{\rm th}\approx0.83~m_ec^2\Gamma_1^{-1}{(A/56)}^{-1/6}$.

The wind driven bubble (or jet) may be filled with photons provided by the CCSN explosion (or shocked jet), since the shocked particles or radioactive nuclei provide x and/or gamma rays, cascading down to optical photons as a result of thermalization.  For example, in CCSNe, the photon temperature of $\approx11~{\rm keV}~{\mathcal E}_{\rm ph, 49}^{1/4}R_{w,10}^{-3/4}$ allows nuclei to survive.  However, disintegration can be caused by high-energy photons from nonthermal electrons accelerated at shocks or by magnetic reconnections.  As in pulsar wind nebulae, we here consider possible effects of nonthermal electrons produced around $R_w$.  The PNS wind is Poynting-dominated after $t_{\rm tr}$, and the magnetic field is $B\sim2.1\times{10}^{8}~{\rm G}~B_{\rm dip,15}P_{-2}^{-1}f_{\rm op,-1}R_{w,10}^{-1}$ from the shock-jump conditions.  The reverse shock is hydrodynamically weak since the thermal energy in the downstream is much smaller than the magnetic energy, but small dissipation can be enough.  The random energy per particle is $0.75\Gamma m_pc^2$ when ions are mainly protons, and the injection Lorentz factor of electrons is $\gamma_{e,i}\approx0.75\epsilon_e\Gamma(m_p/m_e)Y_{e}^{-1}$.  Then, the characteristic synchrotron energy is $\varepsilon_{\rm syn}\sim4.6~{\rm MeV}~\epsilon_{e,-1}^2Y_e^{-2}\Gamma_1^2B_{\rm dip,15}P_{-2}^{-1}f_{\rm op,-1}R_{w,10}^{-1}\gtrsim m_ec^2$, which is higher than the pair-production threshold of $\sim m_ec^2$.  Since the shocked wind or jet is filled with copious photons, synchrotron cascades will be developed and nuclei in the unshocked wind or jet typically interact with boosted $\lesssim2\Gamma m_ec^2$ photons.  For $\varepsilon_{\rm th}<m_ec^2$, using the photon spectrum is $\propto \varepsilon^{-0.5}$ in the fast cooling case, the comoving photon density at $\varepsilon_{\rm th}$ is
\begin{equation}
n_\gamma\sim(1.5\epsilon_e\Gamma^2n_wm_pc^2/\varepsilon_{\rm syn}){(\varepsilon_{\rm th}/\varepsilon_{\rm syn})}^{-0.5}.
\end{equation}
The Thomson optical depth in the PNS wind is estimated to be $\tau_T\approx y_{\pm}n_e\sigma_T(R_w/\Gamma)\simeq0.066y_\pm Y_e\Gamma_{1}^{-2}R_{w,10}^{-1}{(\dot{M}_b/{10}^{-10.5}~M_{\odot}~{\rm s}^{-1})}$, where $y_\pm$ is the possible enhancement factor by pairs that are produced at $R_w$.  Hence, at sufficiently late times, high-energy photons may leave the shocked flow.  Using the effective cross section $\kappa_A\sigma_{A\gamma}\approx1.4\times{10}^{-27}~{\rm cm}^2~{(A/56)}^{1/6}$~\cite{rac96}, the photodisintegration efficiency $f_{A \gamma}\sim\kappa_A\sigma_{A \gamma}(2\Gamma n_\gamma)(R_w/\Gamma)$ is
\begin{eqnarray}
f_{A\gamma}&\sim&66{(A/56)}^{1/6}Y_e\Gamma_1^{1/2}\sigma_3^{-1}\nonumber\\
&\times&B_{\rm dip,15}^{3/2}P_{-2}^{-3/2}f_{\rm op,-1}^{3/2}R_{w,10}^{-1/2}.
\end{eqnarray}
Eq.~(13) suggests that synchrotron photons may disintegrate nuclei and supply neutrons if $\Gamma\approx{\rm min}[\Gamma (R_w),\Gamma(R_{\rm mag})]$ is high enough and $\tau_T$ is so small that high-energy photons can escape, where an additional contribution of neutrinos can be produced.  Generally speaking, the fraction of neutrons ($X_n$) depends on outflow dynamics as well as the initial entropy and $Y_{e0}$.  Note that $f_{A\gamma}$ declines with time, and nuclei would easily survive at $t\gg t_{\rm thin}$.

%%%%%%%%%%%%%%%%%%%%%%%%%%%%%%%%%%%%%%%%%%%%%%%%%%
%%%%%%%%%%%%%%%%%%%%%%%%%%%%%%%%%%%%%%%%%%%%%%%%%%

\section{Neutrino production}
While the wind or jet excavates the stellar material, ions are quickly decelerated at the shock via radiation or collisionless processes.  On the other hand, relativistic neutrons should be decelerated via $np$ collisions, inevitably leading to neutrino production.  If $\Gamma_n<\Gamma$ due to $R_{\rm dec}<R_w,R_{\rm mag}$, the neutron beam propagate in the wind or jet, which can be damped in the shocked flow since the optical depth for the $np$ reaction is $\approx2\Gamma n_w\sigma_{np}R_w\simeq 1.4B_{\rm dip,15}^2P_{-2}^{-2}f_{\rm op,-1}^2R_{w,10}^{-1}\sigma_3^{-1}$.  Note that, even if $\Gamma_n=\Gamma$, their deceleration scale $\sim1/(\Gamma n_w0.5\sigma_{np})$ is longer than $\sim1/(\Gamma n_e\sigma_T)$ and any relevant plasma scales, so neutrino production is expected when ions are stopped by the reverse shock.  At sufficiently late times, neutrons go through the flow, where they are damped in the stellar material (with mass $M_{\rm sn}$) although the hadronic cooling of mesons and mesons can be relevant.  Although the duration of neutron emission is as short as $\sim t_{\rm thin}$, in principle, neutrons leaving the flow inside the stellar material can be depleted until the CCSN ejecta reaches the pionosphere at $\approx6.5\times{10}^{15}~{\rm cm}~{(M_{\rm sn}/10M_{\odot})}^{1/2}$.    
%$f_{np}^{\rm sn} \approx \kappa_{n} \sigma_{np} n_{\rm sn} R_{\rm rs} = {(R_{\rm rs}/R_{\rm pi})}^{-2}$

When all neutrons are damped, half of the neutron kinetic energy goes to neutrinos, so quasithermal neutrinos have a luminosity of
\begin{equation}
{\mathcal L}_\nu\approx0.5(\Gamma_n-1)X_n\dot{M}_b c^2. 
\end{equation} 
For $t_a<t_b$, assuming $V\propto t^{\chi}$ with small $\chi$, we have ${\mathcal L}_\nu\propto t^{-(4-\chi)/3}$ until $t_b$, so ${\mathcal L}_\nu$ and ${\mathcal E}_\nu$ typically decrease with time.  Inelastic collisions occur only when $\Gamma_n\geq1.37$, almost after the PNS wind becomes relativistic at $\sim t_{\rm tr}$.  (Of course, both $R_w$ and $R_{\rm mag}$ have to be larger than $R_{\rm lc}$.)   Hence, the total energy of neutrinos is ${\mathcal E}_\nu\sim0.5 X_n(0.37\dot{M}_b)|_{t_{\rm tr}}c^2t_{\rm tr}$, and we have 
\begin{eqnarray}
{\mathcal E}_\nu^{\rm iso}&\sim&7.8\times{10}^{48}~{\rm erg}~X_{n} B_{\rm dip,15}^{6/5}P_{-2}^{-6/5}f_{\rm op}^{8/5}f_{\rm cen}^{2/5}\nonumber\\
&\times&L_{\nu 0,52}^{2/3}{(\varepsilon_{\nu 0}/15~{\rm MeV})}^{4/3}f_{\rm es}^{2/3}f_b^{-1}, 
\end{eqnarray}
where $f_b$ is the beaming factor and the typical neutrino energy is $E_\nu\approx0.25m_\pi c^2\simeq35~{\rm MeV}$.  If the PNS wind is collimated as in GRB jets, the observed fluence is enhanced by $f_b^{-1}\sim100$, where such neutrinos could be seen as a tail by water-Cherenkov neutrino detectors such as Super-Kamiokande (SK) and a next-generation detector HK.  Note that such jets may be choked rather than successful.  

Since $\Gamma_n$ increases with time, higher-energy neutrinos are produced later, and these are more easily detected.  However, $L_\nu$ and ${\mathcal L}_\nu$ decline abruptly after $t_{\rm thin}$.  For neutrinos originating from intrinsic neutrons, the high-energy contribution is mainly generated at $\sim t_b$, at which we have ${\mathcal E}_\nu\sim0.5X_n(\Gamma_n\dot{M}_b)|_{t_b}c^2t_b$ and Eq.~(11) leads to
\begin{eqnarray}
{\mathcal E}_\nu^{\rm iso}&\sim&2.7\times{10}^{47}~{\rm erg}~X_{n}B_{\rm dip,15}^{26/35}P_{-2}^{-39/25}f_{\rm op,-1}^{46/35}f_{\rm cen}^{4/7}\epsilon_{\rm rec,-2}^{13/35}\nonumber\\
&\times&L_{\nu 0,52}^{20/21}{(\varepsilon_{\nu 0}/15~{\rm MeV})}^{40/21}f_{\rm es}^{20/21}V_{8.5}^{3/7}f_b^{-1}.
\end{eqnarray}
Although Eq.~(16) is valid for neutrons produced via photodisintegration of nuclei, they achieve higher Lorentz factors at $t\gtrsim t_b$ and their contribution at $\sim t_{\rm thin}$ may be relevant.  Using ${\mathcal E}_\nu\sim0.5X_n(\Gamma(R_w)\dot{M}_b)|_{t_{\rm thin}}c^2t_{\rm thin}$,
\begin{eqnarray}
{\mathcal E}_\nu^{\rm iso}&\sim&3.8\times{10}^{46}~{\rm erg}~X_{n}B_{\rm dip,15}^{2/3}P_{-2}^{-1}f_{\rm op,-1}^{4/3}f_{\rm cen}^{2/3}\epsilon_{\rm rec,-2}^{1/3}\nonumber\\
&\times&L_{\nu 0,52}^{10/9}{(\varepsilon_{\nu 0}/15~{\rm MeV})}^{20/9}f_{\rm es}^{10/9}R_{w,10}^{1/3}f_{b}^{-1},
%{(t_{\rm thin}/50~{\rm s})}^{-2/3}
\end{eqnarray}
is obtained for this case (where $t_{\rm thin}=50$~s is used).

For $t_b<t_a$, ${\mathcal L}_\nu\propto\Gamma_n\dot{M}_b\propto t^0$ until $t_a$ and $\Gamma_n$ saturates when $R_{\rm mag}=R_{\rm dec}$.  Then, for both the origins of neutrons, we have ${\mathcal E}_\nu\sim0.5X_n(\Gamma_n\dot{M}_b)|_{t_a}c^2t_a$ and
\begin{eqnarray}
{\mathcal E}_\nu^{\rm iso}&\sim&9.6\times{10}^{47}~{\rm erg}~X_nB_{\rm dip,15}^{34/25}P_{-2}^{-36/25}f_{\rm op,-1}^{44/25}f_{\rm cen}^{2/5}\epsilon_{\rm rec,-2}^{2/25}\nonumber \\
&\times&L_{\nu 0,52}^{2/3}{(\varepsilon_{\nu 0}/15~{\rm MeV})}^{4/3}f_{\rm es}^{2/3}f_{b}^{-1}.
\end{eqnarray}
In either of Eqs.~(16)-(18), the quasithermal neutrino spectrum may extend to $\sim\Gamma_{n,1}$~GeV with the typical energy
\begin{equation}
E_\nu^{\rm qt}\approx0.05\Gamma_{n}m_nc^2\simeq0.47~{\rm GeV}~\Gamma_{n,1}
\end{equation}
for $\Gamma_n\gg1$.  Such GeV neutrinos are good targets for PINGU, a planned low-energy extension of IceCube, as well as HK. 

In addition, when the ion and neutron flows are coupled up to the shock radius $R_w$, the NPC acceleration mechanism will work, and the typical energy of boosted nucleons is~\cite{npc}  
\begin{equation}
E_\nu^{\rm NPC}\approx0.1{(\kappa_N\Gamma_{n})}^3m_nc^2\simeq12~{\rm GeV}~\Gamma_{n,1}^3,
\end{equation}
where $\kappa_N\approx0.5$.  The efficiency of the NPC acceleration can be $\gtrsim10$\% of the neutron-flow energy~\cite{npc}, although detailed studies are left as future work, it can enhance the detectability of multi-GeV neutrinos especially if other nonthermal particle production is inefficient.  Ions may also be accelerated by the shock acceleration mechanism even at subphotospheres, leading to much higher-energy neutrinos, as often considered in the context of GRBs~\cite{mur08}.  Although details are uncertain, the combination of the shock-driven magnetic reconnection and shock acceleration at the termination shock~\cite{ss11} may be relevant. 

In the standard pulsar phase ($t\gg t_{\rm thin}$), $\dot{M}\approx\dot{M}_{\rm GJ}$ and we expect few intrinsic neutrons.  The wind will be dominated by electron-positron pairs, though nuclei may be stripped from the surface.  Neutrons may be supplied by photodisintegration.  
However, as the target photon density is low, photodisintegration of thermal nuclei and quasithermal neutrino production are likely to become inefficient. 
On the other hand, possible nonthermal ions can efficiently interact with softer nonthermal photons, leading to high-energy neutrinos.  

%%%%%%%%%%%%%%%%%%%%%%%%%%%%%%%%%%%%%%%%%%%%%%%%%%
%%%%%%%%%%%%%%%%%%%%%%%%%%%%%%%%%%%%%%%%%%%%%%%%%%

\section{Quasithermal neutrino detection}
From Eqs.~(15)-(18), the energy fluence of quasithermal neutrinos per flavor is $E_\nu^2\phi_\nu\simeq28~{\rm erg}~{\rm cm}^{-2}~{\mathcal E}_{\nu,48}^{\rm iso} {(D/10~{\rm kpc})}^{-2}$. 
%\begin{equation}
%E_\nu^2 \phi_\nu \approx \frac{1}{4 \pi D^2} \frac{{\mathcal E}_\nu^{\rm iso}}{3} \simeq 2.7~{\rm erg}~{\rm cm}^{-2}~{\mathcal E}_{\nu,47}^{\rm iso} {\left( \frac{D}%{10~{\rm kpc}} \right)}^{-2}.  
%\end{equation}

PINGU has sensitivity to $1-10$~GeV neutrinos with an effective area of $\sim4\times{10}^{-3}~{\rm cm}^2$ for $\nu_e+\bar{\nu}_e$ and $\sim2\times{10}^{-3}~{\rm cm}^2$ for $\nu_\mu+\bar{\nu}_\mu$, respectively, at $\sim1$~GeV.  So, $\sim100~{\mathcal E}_{\nu,48}^{\rm iso}$ events are expected for a CCSN at 10~kpc.  In the case of choked jets the energy fluence may be enhanced by $f_b^{-1}\sim100$, and the detection of GeV neutrinos becomes possible even for extragalactic CCSNe out to $\sim1$~Mpc.      

HK has a fiducial volume of 0.56~Mt, so the effective numbers of free protons and bound nucleons in oxygen are $3.7\times{10}^{34}$ and $3.0\times{10}^{35}$, respectively.  The neutrino-nucleon cross section for the charged-current interaction at 1~GeV is $\sim0.6\times {10}^{-38}~{\rm cm}^2$ (averaged over $\nu$ and $\bar{\nu}$), so the effective area is $\sim2\times{10}^{-3}~{\rm cm}^2$.  Hence, we may expect $\sim70~{\mathcal E}_{\nu,48}^{\rm iso}$ events for a CCSN at 10~kpc.  In addition, HK could also allow us to see $\sim10-100$~MeV neutrinos through the $\bar{\nu}_ep\rightarrow e^+n$ channel.  However,  detection of these lower energy neutrinos would be more difficult because of the smaller cross sections at lower energies and because the signal may be buried in the exponential tail of thermal MeV neutrinos from the PNS.   

To see the signal, it is crucial to reduce backgrounds using space and time coincidence.  The obvious background is the atmospheric neutrino background (ANB).  The ANB at GeV is $\approx1.3\times{10}^{-2}~{\rm GeV}~{\rm cm}^{-2}~{\rm s}^{-1}~{\rm sr}^{-1}$ for $\nu_e+\bar{\nu}_e$ and $\approx2.6\times{10}^{-2}~{\rm GeV}~{\rm cm}^{-2}~{\rm s}^{-1}~{\rm sr}^{-1}$ for $\nu_\mu+\bar{\nu}_\mu$, respectively~\cite{hon+11}.  We may take the time window of $t_{\rm thin}\sim10-100$~s after the explosion time that is measurable with MeV neutrinos and/or gravitational waves~\cite{gw}.   The localization is possible by follow-up observations at x-ray, optical, and infrared bands.  In addition, radio observations may be useful for this purpose.  Although radio supernovae have been observed only for a fraction of CCSNe and most of them seem to simply arise from the existence of dense circumstellar martial, some CCSNe such as SN 1986J suggest possible activities of pulsars embedded in CCSN ejecta~\cite{radiosne}.  Even though they are observed with much longer time scales compared to the duration of neutrino emission, future coincident observations would be useful.  The angular resolution of PINGU is expected to be $\sim2-20$~deg~\cite{pingu}, which is typically much larger than angular resolutions of photon observations.  But, the ANB in this angular window and the time window of $t_{\rm thin}$, which is $\sim2\times{10}^{-3}~{\rm erg}~{\rm cm}^{-2}$, is small enough.  

It would be difficult to see the diffuse neutrino background from quasithermal neutrinos discussed here.  Since the released neutrino energy per explosion is not large, other contributions such as thermal neutrinos from CCSNe and nonthermal neutrinos produced by cosmic rays in star-forming galaxies will be more relevant. 

%%%%%%%%%%%%%%%%%%%%%%%%%%%%%%%%%%%%%%%%%%%%%%%%%%
%%%%%%%%%%%%%%%%%%%%%%%%%%%%%%%%%%%%%%%%%%%%%%%%%%

\section{Implications and discussions}
Neutron-loaded relativistic winds emanate from rotating and magnetized PNSs during their $\sim10-100$~s cooling epoch.  In this work, we show that relativistic neutrons produce $\sim0.1-1$~GeV or even $\gtrsim1$~GeV neutrinos via the $np$ reaction as the wind or jet interacts with the surrounding stellar material.  Such a role for neutrons in generating neutrino emission has been studied in the context of GRBs.  Here, we have considered PNSs in general, including quasispherical winds and choked jets.  Interestingly, the production of GeV neutrinos, does not rely on uncertain cosmic-ray ion acceleration mechanisms that lead to nonthermal neutrinos, and instead relies primarily on magnetocentrifugal acceleration.  We also pointed out that, however, even higher-energy neutrinos may be additionally produced via some particle acceleration mechanism.  The shock acceleration mechanism is the most popular process, which is directly observed in data of solar winds~\cite{solar} and widely accepted in the context of supernova remnants.  Since PNS winds may still be magnetically dominated at the termination shock, it is not clear what is the most important process to generate nonthermal particles, and examples of particle acceleration mechanisms include not only the shock acceleration at the termination shock but also the magnetic reconnection and wakefield-like acceleration in the PNS wind, which could generate ultrahigh-energy cosmic rays and result in PeV-EeV neutrinos via interactions with the stellar material and CCSN photons~\cite{gt87,mur+09}.  The physical conditions producing such nonthermal neutrinos, which may occur only at $t\gg t_{\rm thin}$ (where $\dot{M}\approx\dot{M}_{\rm GJ}$), is completely different from the case considered in this work.  

Effects related to neutrino propagation in the stellar material may be important.  The optical depth for the $\nu N$ interaction is small.  The quasispherical wind interacts with the shocked stellar material, and we have $\sim0.1(M_{\rm sn}/10M_\odot)R_{\rm sn,10}^{-2}$ at GeV, where $R_{\rm sn}$ is the CCSN ejecta radius.  The jet may interact with the preshocked envelope, and we have only $\sim{10}^{-4}$ at GeV if $\rho\sim1~{\rm g}~{\rm cm}^{-3}$ at ${10}^{10.5}$~cm.  Studying neutrino oscillations in the MeV-GeV range would be much more interesting.  For example, the matter effect is relevant since the resonance happens at $\approx0.22~{\rm g}~{\rm cm}^{-3}~{(E_\nu/\rm GeV)}^{-1}(\Delta m^2/7.59\times{10}^{-5}~{\rm eV}^2)$, and we can use it to probe the density profile or check neutrino properties given the progenitor structure~(c.f.~\cite{neuosc}).

Although there appears to be a number of uncertain parameters governing the neutrino emission from the processes described in this work, the qualitative picture is simple.  The parameters can be basically classified into those related to $\dot{M}_b$ and those related to $\Gamma_n$.  Regarding $\dot{M}_b$, one sees that $L_\nu$, $\varepsilon_\nu$ and $f_{\rm es}$ are determined by the PNS physics, and are relatively well-known from earlier theoretical work on PNS winds and neutron star cooling, and from the neutrinos detected directly from SN 1987A.  For this reason, the values will not differ largely from our fiducial values.  On the other hand, $f_{\rm op}$ and $f_{\rm cen}$ can be regarded as subparameters, and although their detailed time evolution is uncertain in the PNS context, their physics is reasonably understood.  The most relevant parameters are the physical parameters of the PNS that determine the time-evolution of the PNS wind magnetization and its power: $B_{\rm dip}$ and $P$.  These directly determine $\sigma$, which determines the Lorentz factor of the neutrons, $\Gamma_n$.  
Throughout this work, we consider PNSs, where $\dot{M}_\nu$ is given by Eq.~(2).  The similar discussion can be made in the case of black hole formation, given that $\dot{M}_\nu$ is calculated for an accretion disk.  

Our results are general as long as $\sigma$ is high enough that the flow becomes relativistic at times before the PNS becomes optically-thin to neutrinos $t_{\rm thin}$; for magnetar strength dipole fields and rotation periods less than $\sim10$~ms this criterion is fulfilled.  Indeed, our results imply that for PNSs with $B_{\rm dip}\gtrsim10^{14}$~G and $P\sim1-10$~ms future neutrino telescopes such as PINGU and HK may detect $\sim10-100$ neutrino events of energy $\sim0.1-1$~GeV from the next magnetar-producing Galactic CCSN.  Unfortunately, not all PNSs are expected to be born with such high $B_{\rm dip}$ and short $P$.  As a reference value, if we use the magnetar birth rate which is $\gtrsim10$\% of the CCSN rate~\cite{magrev}, the chance of seeing a Galactic event is very small.  Even so, nonmagnetars may also be detected thanks to particle acceleration and in principle even extragalactic CCSNe may be detected if the wind is collimated, forming choked jets.  In the latter case, we expect $0.02-0.05f_b~{\rm yr}^{-1}$ for the birth of magnetars within 5~Mpc~\cite{mur+09}.  Given the fact that the ANB is reduced by detections that are reasonably coincident with follow-up observations at x-ray, optical, infrared, and radio bands, stacking analyses for nearby CCSNe would be helpful. 

The detection of quasithermal neutrinos would strongly suggest the existence of relativistic neutron outflows, which cannot be probed by photon observations.  Their detection is possible by next-generation neutrino telescopes such as PINGU and HK, which should provide us with precious insights into magnetic acceleration mechanisms, the physics of PNSs, and clues to nucleosynthesis.  Or, with good knowledge of $\dot{M}_b$, nondetections of GeV neutrinos can potentially limit $B_{\rm dip}$ and $P$ via constraints on $\Gamma_n$.  Multi-wavelength studies at radio, optical and x-rays are also relevant to test our picture and constrain $B_{\rm dip}$ and $P$ by searching for long-term energy injection by PNSs.  Rapidly-rotating and strongly-magnetized PNSs have been proposed as the central engine of successful jets leading to GRBs in compact stellar progenitors.  A larger fraction of PNSs would have less extreme $B_{\rm dip}$ and $P$, leading to failed GRBs, hypernovae and perhaps super-luminous supernovae.  Detecting the GeV neutrinos would also be useful in revealing this CCSN-GRB connection.  

Very recently, the Gton neutrino detector, IceCube, reported the likely discovery of astrophysical high-energy neutrinos~\cite{PeVneu}.  The energy of neutrino-induced showers lies in the 30~TeV-1~PeV range.  In the context of PNSs, as pointed out in this work, such high-energy neutrinos may be produced via some particle acceleration mechanism in the wind or around the termination shock.  However, IceCube is not sufficient for hunting much more guaranteed quasithermal neutrinos since it was built mainly for detecting neutrinos above TeV energies.  Present water-Cherenkov detectors such as SK seem too small to detect GeV neutrinos from astrophysical sources.  Thus, we encourage having sufficiently large neutrino detectors that can fill the gap between MeV and TeV energies, which include DeepCore and PINGU as well as future Mton neutrino detectors like HK.

%%%%%%%%%%%%%%%%%%%%%%%%%%%%%%%%%%%%%%%%%%%%%%%%%%
%%%%%%%%%%%%%%%%%%%%%%%%%%%%%%%%%%%%%%%%%%%%%%%%%%

\medskip
\acknowledgements
We thank John Beacom, Kumiko Kotera, Peter M\'esz\'aros, Brian Metzger, especially Masayuki Nakahata and Carsten Rott for helpful comments.  We acknowledge support by the CCAPP workshop, Revealing Deaths of Massive Stars with GeV-TeV Neutrinos, and NASA through Hubble Fellowship, Grant No. 51310.01 awarded by the STScI, which is operated by the Association of Universities for Research in Astronomy, Inc., for NASA, under Contract No. NAS 5-26555 (K. M.).

%%%%%%%%%%%%%%%%%%%%%%%%%%%%%%%%%%%%%%%%%%%%%%%%%%
%%%%%%%%%%%%%%%%%%%%%%%%%%%%%%%%%%%%%%%%%%%%%%%%%%


\begin{thebibliography}{99}
\bibitem{bl86}
A. Burrows and J.~M. Lattimer, Astrophys. J. {\bf 307}, 178 (1986).
\bibitem{pon+99}
J.~A. Pons {\it et al.}, Astrophys. J. {\bf 513},780 (1999). 
\bibitem{woo+94}
S.~E. Woosley {\it et al.}, Astrophys. J. {\bf 433}, 209 (1994).  
\bibitem{qw96}
Y.-Z. Qian and S.~E. Woosley, Astrophys. J. {\bf 471}, 331 (1996).
\bibitem{tho+04}
T.~A. Thompson, P. Chang, and E. Quataert, Astrophys. J. {\bf 611}, 380 (2004). 
\bibitem{met+11}
B.~D. Metzger {\it et al.}, Mon. Not. R. Astron. Soc. {\bf 413}, 2031 (2011).
\bibitem{mic69}
F.~C. Michel,  Astrophys. J. {\bf 158}, 727 (1969). 

\bibitem{rob+10}
L.~F.~Roberts, S.~E. Woosley, and R.~D. Hoffman, Astrophys. J. {\bf 722}, 954 (2010).
\bibitem{met+11b}
B.~D. Metzger, D. Giannios, and S. Horiuchi, Mon. Not. R. Astron. Soc. {\bf 415}, 2495 (2011).
\bibitem{gj70}
P. Goldreich and W.~H. Julian, Astrophys. J. {\bf 160}, 971 (1970). 
\bibitem{kom11}
S.~S. Komissarov, Mem. della Soc. Astron. Italiana {\bf 82}, 95 (2011).
\bibitem{neutron}
E.~V. Derishev, V.~V. Kocharovsky, and V.~V. Kocharovsky, Astrophys. J. {\bf 521}, 640 (1999);
J.~N. Bahcall and P. M\'esz\'aros, Phys. Rev. Lett. {\bf 85}, 1362 (2000);
P. M{\'e}sz{\'a}ros and M.~J. Rees, Astrophys. J. {\bf 733}, L40 (2011).

\bibitem{magrev}
P.~M. Woods and C. Thompson, Compact Stellar X-ray Sources, edited by W.~H.~G. Lewin and M. van der Klis, Cambridge Astrophysics Series Vol. 39 (2006), p. 547; S. Mereghetti, Astron. Astrophys. Rev. {\bf 15}, 225 (2008).
\bibitem{dt92}
R.~C. Duncan and C. Thompson, Astrophys. J. {\bf 392}, 9 (1992); 
C. Thompson and R.C. Duncan, Astrophys. J. {\bf 408}, 194 (1993). 
\bibitem{buc+07}
N. Bucciantini {\it et al.}, Mon. Not. R. Astron. Soc. {\bf 380}, 1541 (2007); 
{\bf 396}, 2038 (2009).
\bibitem{kb07}
S.~S. Komissarov and M.~V. Barkov, Mon. Not. R. Astron. Soc. {\bf 382}, 1029 (2007).
\bibitem{uso94}
V. Usov, Nature {\bf 357}, 472 (1994). 
\bibitem{ccsnjet}
E. Berger, S.~R. Kulkarni, D.~A. Frail, and A.~M. Soderberg, Astrophys. J. {\bf 599}, 408 (2003).
A.~M. Soderberg, E. Nakar, E. Berger, and S.~R. Kulkarni, Astrophys. J. {\bf 638}, (2006).
\bibitem{sod+06}
A.~M. Soderberg {\it et al.}, Nature (London) {\bf 442}, 1014 (2006).
\bibitem{sod+10}
A.~M. Soderberg {\it et al.}, Nature (London) {\bf 463}, 513 (2010).
\bibitem{wm03}
E. Waxman and P. M\'esz\'aros, Astrophys. J. {\bf 584}, 390 (2003). 
\bibitem{whe+00}
J.~C. Wheeler, I. Yi, P. H{\"o}flich, and L. Wang, Astrophys. J. {\bf 537}, 810 (2000); 
D. Kasen and L. Bildsten, Astrophys. J. {\bf 717}, 245 (2010). 
\bibitem{moscow}
G.~S. Bisnovatyi-Kogan, Astron. Zh. {\bf 47}, 813 (1970);
G.~S. Bisnovatyi-Kogan, Yu.~P. Popov, and A.~A. Samochin, Astrophys. Space Sci. {\bf 41}, 287 (1976);
N.~V. Ardeljan, G.~S. Bisnovatyi-Kogan, and S.~G. Moiseenko, Mon. Not. R. Astron. Soc. {\bf 359}, 333 (2005); 
S.~G. Moiseenko, G.~S. Bisnovatyi-Kogan, and N.~V. Ardeljan, Mon. Not. R. Astron. Soc. {\bf 370}, 501 (2006).
\bibitem{mh79}
E. Mueller and W. Hillebrandt, Astron. Astrophys. {\bf 80}, 147 (1979).

\bibitem{npc}
%E.~V. Derishev, F.~A. Aharonian, V.~V. Kocharovsky, and V.~V. Kocharovsky, Phys. Rev. D {\bf 68}, 043003 (2003);
K. Murase, K. Kashiyama, and P. M\'esz\'aros, Phys. Rev. Lett. {\bf 111}, 131102 (2013); 
K. Kashiyama, K. Murase, and P. M\'esz\'aros, Phys. Rev. Lett. {\bf 111}, 131103 (2013).
\bibitem{pingu}
D.~J. Koskinen, Mod. Phys. Lett. A {\bf 26}, 2899 (2011); 
K. Clark and D.~F. Cowen, Nucl. Phys. B Proc. Suppl. {\bf 00}, 1 (2012).
\bibitem{hk}
K. Abe {\it et al.}, arXiv:1109.3262 (2011). 

\bibitem{tho+05}
T.~A. Thompson, E. Quataert, and A. Burrows, Astrophys. J. {\bf 620}, 861 (2005). 
\bibitem{met+08}
B.~D. Metzger, T.~A. Thompson, and E. Quataert, Astrophys. J. {\bf 676}, 1130 (2008).
\bibitem{dq04}
H. Duan and Y.-Z. Qian, Phys. Rev. D {\bf 69}, 123004 (2004).

\bibitem{gj69}
P. Goldreich and W.~H. Julian, Astrophys. J. {\bf 157}, 869 (1969). 
%\bibitem{wlw93}
%S.~E. Woosley, N. Langer, and T.~A. Weaver, Astrophys. J. {\bf 411}, 823 (1993).

\bibitem{spr+03}
H.~C. Spruit, F. Daigne, and G. Drenkahn, Astrophys. Astron. {\bf 369}, 694 (2001); 
G. Drenkann, Astrophys. Astron. {\bf 387}, 714 (2002).
\bibitem{ber13}
A. Beresnyak, arXiv:1301.7424.

\bibitem{hor+12}
S. Horiuchi {\it et al.}, Astrophys. J. {\bf 753}, 69 (2012).
\bibitem{rac96}
J.~P. Rachen, Ph. D Thesis (1996).
\bibitem{mur08}
K. Murase, Phys. Rev. D {\bf 78}, 101302(R) (2008); 
X.-Y. Wang and Z.-G. Dai, Astrophys. J. {\bf 691}, L67 (2009).
\bibitem{ss11} 
L. Sironi and A. Spitkovsky, Astrophys. J. {\bf 741}, 39 (2011).

\bibitem{hon+11}
M. Honda, T. Kajita, K. Kasahara and S. Midorikawa, Phys. Rev. D {\bf 83}, 123001 (2011).
\bibitem{gw}
I. Bartos, P. Brady, and S. M\'arka, Classical and Quantum Gravity {\bf 30}, 123001 (2013).
\bibitem{radiosne}
K.~W. Weiler {\it et al.}, Astrophys. J. {\bf 301}, 790 (1986);
K.~W. Weiler, N. Panagia, and R.~A. Sramek, Astrophys. J. {\bf 364}, 611 (1990).

\bibitem{solar}
G.~P. Zank, W.~K.~M. Rice, and C.~C. Wu, J. Geophys. Res. {\bf 105}, 25079 (2000); 
J. Giacalone, J.~F. Drake, and J.~R. Jokipii, Space Sci. Rev. {\bf 173}, 283 (2012).

\bibitem{mur+09}
K. Murase, P. M\'esz\'aros, and B. Zhang, Phys. Rev. D {\bf 79}, 103001 (2009).
\bibitem{gt87}
T.~K. Gaisser and T. Stanev, Phys. Rev. Lett. {\bf 58}, 1695 (1987); 
W. Bednarek and R.~J. Protheroe, Phys. Rev. Lett. {\bf 79}, 2616 (1997);  
J.~H. Beall and W. Bednarek, Astrophys. J. {\bf 569}, 343 (2002).
%\bibitem{aro03}
%J. Arons, Astrophys. J. {\bf 589}, 871 (2003). 

\bibitem{neuosc}
R.~C. Schirato and G.~M. Fuller,
%``Connection between supernova shocks, flavor transformation, and the neutrino signal,''
arXiv:astro-ph/0205390;
%%CITATION = ASTRO-PH/0205390;%%
K. Takahashi, K. Sato, A. Burrows, and T.~A. Thompson, Phys. Rev. D {\bf 68}, 113009 (2003).

\bibitem{PeVneu}
M. Aartsen {\it et al.}, Phys. Rev. Lett. {\bf 111}, 021103 (2013);
M. Aartsen {\it et al.}, Science {\bf 342}, 1242856 (2013).
\end{thebibliography}
\end{document}